\documentclass[useAMS,usenatbib]{mn2e}
\usepackage{amssymb,amsmath,rotating,graphicx}
\usepackage[]{natbib}

\title[3-D Relativistic SPH]{3-D Relativistic SPH}
\author[S. Muir and J.J. Monaghan]{S. Muir$^{1}$\thanks{E-mail:
Stuart.Muir@maths.monash.edu.au} and J.J. Monaghan$^{1}$\footnotemark[1]\thanks{E-mail:Joe.Monaghan@sci.monash.edu.au}\\
$^{1}$Centre for Stellar and Planetary Astrophysics, \\
School of Mathematical Sciences,\\
 Monash University,\\
Wellington Rd. Clayton,\\
Victoria, AUS 3800}
\begin{document}

\date{Accepted ******. Received ******* ; in original form 2003 November 25}

\pagerange{\pageref{firstpage}--\pageref{lastpage}} \pubyear{2002}

\maketitle

\label{firstpage}

\begin{abstract}
In this paper we present the equations and some basic applications of relativistic SPH. The equations are generated under the assumption that there are no interaction terms between the fluid modelled and the background space-time, i.e. we are neglecting the perturbations to the metric produced by the modelled fluid. This corresponds to a stationary metric, which for this work, we further limit to be static as well. The equations use a new signal velocity term and artificial viscosity to smear out the effects of strong shocks and are tested here against Newtonian and relativistic shocks.
 \end{abstract}

\section{Introduction}

The advent of x-ray telescopes with resolutions of \emph{Chandra} and \emph{RXTE} has opened the window into high energy astrophysics. Although we now have increasingly complex and detailed data sets in these higher wavebands, the theory and our ability to understand these images is falling far behind. The reason for this is the complexities 
involved in solving the equations of relativistic fluid dynamics, an essential aspect of high energy astrophysical phenomena.  In general, the simulation of these systems requires the solution of the General Relativistic field equations as well as the solution of the equations of fluid dynamics. This is a complex and difficult computational problem which has only been met with limited success.(See \citet{livrev} for a detailed review) 

However, there is a range of problems where the metric can be specified independently of the fluid motion  and it is to this class of problems that the present paper is directed.  Of these problems the simplest involves motion in the flat space of special relativity and a typical example is a relativistic jet (see \citet{livingreview}).
 Even under these simplified conditions, these types of flow are notorious for the computational complexities that  arise when one tries to model them. This is basically due to the strong shocks and the narrow, dense structures that form easily in the supersonic flow.

One approach to these problems is to set up a finite difference scheme and make use of the exact Riemann solutions for an ideal gas.  This approach has been very successful in the hands of \citet{livingreview} and others (\citet{schneider93} for example)
 who have obtained very accurate results for one and two dimensional shock wave problems involving shock tubes and jets.  A disadvantage of these methods is that the Riemann solution must be known and for complex problems this may be difficult to obtain.  For example, in a relativistic collision of two streams new particles may be created. The current relativistic Riemann solutions do not include such processes. Another example is the fluid dynamical model of  high energy nuclei collisions where the equation of state is a complicated function of the densities and energies of the colliding fluid.  No relativistic Riemann solutions have been obtained for these equations of state,  so that approximate Riemann solvers, of doubtful accuracy and stability must be used.

An alternative approach is to avoid the discontinuities by smoothing them over some length scale, usually by an artificial viscosity term, and evolving the hydrodynamical equations numerically. If this is done on an Eulerian grid, the questions of resolving the fine structures mentioned above arise, as do the problems of the massive grid distortions which will be experienced in the vicinity of compact objects. These problems can be avoided by using Lagrangian particle methods. The method which forms the basis of this paper, is the particle method SPH (for a review see \citet{joe92} and for applications to one dimensional shock problems see \citet{chowandjoe}). Despite the difficulties with modelling some boundary conditions, we believe that an entirely grid free method is the best solution to modelling these kinds of flows.

 The advantage of SPH is that it handles shocks by prescribing an appropriate artificial dissipation rather than by the use of exact Riemann solutions.  Although such artificial dissipation can lead to greater dissipation than is the case for Riemann solver methods, the fact that the SPH resolution adjusts for changes in density compensates because the dissipation terms are proportional to the resolution.  As a result, one dimensional tests show that SPH could give results with similar accuracy to either the Riemann methods or the HLLE method \citep{chowandjoe}.  In this paper we describe our algorithm in detail and concentrate our tests on shock tube problems in flat space.  Application to both jets and disks and high energy ion collisions will be given elsewhere.

Throughout the paper we use the Einstein summation convention, where Greek indices ($\mu,\nu$) indicate four vector components (0,1,2,3), and Latin ($i,j$) signify spatial summations only (1,2,3). We work in geometric units, where $c=G=k=1$. And finally, we note that throughout our SPH equations we tag quantities  associated with particles by the indices $a$ and $b$.

\section{The Relativistic Gas Equations}
We assume the gas consists of identical baryons with rest mass $m_0$ moving with four velocity $U^{\mu}$ in a  space with metric coefficients $g^{\mu \nu}$. We can then completely specify the state of the fluid by defining this four velocity, the baryon number density ($n$), the internal energy ($\epsilon$) and the isotropic pressure, ($P$). The  energy momentum tensor can be written

\begin{equation}
T^{\mu\nu} = (n + n \epsilon +P)U^\mu U^\nu +Pg^{\mu\nu}
\end{equation}
These quantities are measured in the frame co-moving with the element of fluid.   We assume that an equation of state specifying $P$ in terms of the $\epsilon$ and $n$ is available together with a prescription for determining the composition for given energy and density.  In the present paper we assume the equation of state is for an ideal gas of relativistic baryons however, any equation of state can be applied. 

We are assuming here that the gravitational perturbation effects of the modelled fluid can be ignored. This translates to a background metric which is stationary. In order to decouple the time and spacelike integrations and interpolations, we use the 3+1 splitting of the space-time first done by \citet{york}. This yields metrics of the form

\begin{equation}
ds^2= -(\alpha^2-\beta^i\beta_i)dt^2 +2\beta_idx^idt+\eta_{ij}dx^idx^j\end{equation}

In this version of the equations, we further impose staticity upon the metric through enforcing that the shift vector ($\beta^i$) is zero. The term $\alpha$ is the lapse function and is used to define the distance between the foliated, space-like hypersurfaces. 

We can then start by enforcing the conservation of particles,
\begin{equation}
(nU^\mu)_{;\mu}=0\label{massflux}\end{equation}
and that the divergence of the stress-energy tensor is also zero,

\begin{equation}
T^{\mu\nu}_{\quad;\nu}=0 \label{divT}\end{equation}

From Equation (\ref{massflux})  
\begin{equation}
\partial_tN=-\partial_i(Nv^i),\end{equation}
where $N$ number density measured in the laboratory frame and is given by $N=\gamma n$ and \begin{equation}\gamma=\frac{1}{\sqrt{1-\frac{\eta_{ij}v^iv^j}{\alpha^2}}}\end{equation}

Equation \ref{divT} can be projected in directions parallel and orthogonal to the four-velocity $U^\mu$. The orthogonal components are given by,
\begin{equation}
v^\nu(hU_i)_{,\nu}=\frac{\alpha}{N\gamma}(\frac{1}{2}NhU^\sigma U^\nu g_{\sigma\nu,i}-\partial_iP)
\end{equation}
where 
\begin{equation}
h=1+\epsilon+\frac{P}{n}\end{equation}
and is the relativistic enthalpy. We then define the momentum density $Q^i$ to be 
\begin{align}
Q^i :=& \frac{\sqrt{-g}N}{\alpha}hU^i\notag\\
=&\sqrt{\eta} v^i \gamma^2 (n + n\epsilon + P),\label{momdef}
\end{align}
Here $v^i$ is the usual transport velocity $dx^i/dt$, related to the four velocity through \begin{equation}U^\mu=\frac{\gamma}{\alpha}v^\mu.\end{equation} and $\sqrt{\eta}$ is the determinate of the spatial part of the metric, and can be related to the full metric's determinant through \begin{equation} \sqrt{-g}=\alpha\sqrt{\eta}\end{equation}

The parallel component reduces to

\begin{equation}
\partial_t E + \partial_j( E -g^{00}P)v^j = 0,
\end{equation}
where
\begin{equation}
E = \sqrt{\eta}(n + n\epsilon + P)\gamma^2 - P.\label{energydef}
\end{equation}

The above equations are written in terms of momentum and energy per unit volume.  In order to set up SPH equations we need momentum and energy per baryon. However, due to the fact that we are in curved space, the spatial metric determinant $\sqrt{\eta}$ needs to be incorporated into our spatial integrations.
 
The momentum per particle ${\bf q}$ and the energy per particle $e$ then become \begin{equation}
{\bf q} = \frac{\bf Q}{\sqrt{\eta}N} = {\bf v } \gamma (1+\epsilon+ \frac{P}{n}),
\end{equation}
and
\begin{equation}
 e = \frac{E}{\sqrt{\eta}N} = \gamma(1+\epsilon + \frac{P}{n} ) - \frac{P}{\sqrt{\eta}N},
\end{equation}

Using these definitions we find, using the Lagrangian time derivative  
\begin{equation}
  \frac{d}{dt} = \partial_t + v^j\partial_j,
\end{equation}
that the momentum and energy equations become
\begin{equation}
\frac{d q^i }{dt} = - \frac{1}{N^*}((\sqrt{-g}P)_{,i}-\frac{\sqrt{-g}}{2}T^{\mu\nu}g_{\mu\nu,i})\label{momeqn}
\end{equation}
and
\begin{equation}
\frac{d e}{dt} =  \frac{1}{N^*}((Pg^{00}v^i)_{,i})\label{eneqn}
\end{equation}
with the continuity equation
\begin{equation}
\frac{d N^*}{dt}=-N^* v^i_{,i}\end{equation}
In order to express these variables, we have transformed to the number density $N^*$ ($D^*$ in \citet{siegler99}), which incorporates the volume element through $N^*=\sqrt{\eta}N=\sqrt{\eta}\gamma n$.

 Making this transform alleviates the difficiulties experienced by Laguna et al. (1993) with the approximations to spatial integrals in curved space. It allows us to use the flatspace kernels of traditional SPH (see Section \ref{spheqns}).
 It has also been shown that $N^*$ satisfies the continuity equations, whereas using $N$ results in extra terms being required \citep{siegler99}.
Note also that there are no time derivatives of hydrodynamic variables on the right hand side, making the conservation equations more stable than those of \citet{laguna93} \citep{norman83}.

Particles are moved according to
\begin{equation}
\frac{d {\bf r} }{dt} = {\bf v}
\end{equation}
These equations are identical in form to the non relativistic equations and they can be solved in the same way.

\subsection{SPH relativistic equations\label{spheqns} }

The baryon gas is replaced by a set of SPH particles with the number of baryons in SPH particle $a$ denoted by $\nu_a$.  We use $a$ and $b$ as indices which should not be confused with vector or tensor quantities. Care is taken in the discretization process to ensure that the terms are symmetric in $a$ and $b$, which not only increases calculation efficiency, but helps to ensure conservation of linear and angular momentum.

Smoothed particle hydrodynamics is based upon the approximation that

\begin{equation}<A(x)>=\int A({\bf x'})W(|{\bf x'}-{\bf x}|,h)d{\bf x'},\label{Aint}\end{equation} where $A$ is any field variable, $<A>$ is the approximation to $A$ and $W$ is some kernel, normalised such that \begin{equation}\int W({\bf x})dV=1.\label{wint}\end{equation}
If we are in curved space, we can assume the same relations, except with the addition of $\sqrt{\eta}$, in that
\begin{equation}<A(x)>=\int A({\bf x'})\bar{W}(|{\bf x'}-{\bf x}|,h)\sqrt{\eta}d{\bf x'}\label{Abarint}\end{equation} and
\begin{equation}\int \bar{W}({\bf x})\sqrt{\eta}dV=1. \label{wbarint}\end{equation}

If we then make the assumption that the $\bar{W}$ kernel can be described as some spatially varying function times the usual flatspace kernel, $W$,
\begin{equation}
\bar{W}=f({\bf x})W\end{equation}

\citep{laguna93}, then equation (\ref{wbarint}) becomes
\begin{equation}\int W({\bf x})f({\bf x})\sqrt{\eta}dV=1.\end{equation}
which, if we compare to equation (\ref{wint}) we can see that 

\begin{equation}
f({\bf x})=\frac{1}{\sqrt{\eta}}.\end{equation}

Equation (\ref{Abarint}) then becomes 
\begin{align}
<A(x)>&=\int A({\bf x'})\frac{{W}(|{\bf x'}-{\bf x|},h)}{\sqrt{\eta}}\sqrt{\eta}d{\bf x'}\notag \\
&\simeq \sum_b \nu_b \frac{A_b}{N^*_b}W(|{\bf x}-{\bf x_b}|,h)\label{sph1}\end{align}
where $\nu_b$ is the number of baryons depicted by particle $b$, and given by
\begin{equation}
\nu_b=\sqrt{\eta}N \Delta V\end{equation}

Note how equation (\ref{sph1}) is identical in form to traditional SPH, allowing us to use the usual SPH methods.

 The momentum, continuity and energy equations take the form:\\
Conservation of Momentum:
\begin{align}
\frac{d  q^i_a }{dt} = &-\sum_b\nu_b \left ( \frac{\sqrt{-g_a}P_a}{N_a^{*2}} + \frac{\sqrt{-g_b}P_b}{N_b^{*2}} +\Pi_{ab}  \right ) \nabla^a_i W_{ab}+\notag\\&+(\frac{\sqrt{-g}}{2N^*}T^{\sigma\nu}g_{\sigma\nu,i})_a
\end{align}
 Note: this equation is the same as that generated by \citet{price}, although they used a variational principle.

Conservation of Energy:
\begin{equation}
\frac{d e_a}{dt} = - \sum_b \nu_b \left ( \frac{P_a g^{00}_a{\bf v_a}}{N_a^{*2}} + \frac{P_bg^{00}_b{ \bf v}_b}{N_b^{*2}}  + \Omega_{ab}  \right ) \cdot  \nabla_a W_{ab},
\end{equation}

and Conservation of Baryon Number:
\begin{equation}
\frac{dN^*_a}{dt} = \sum_b \nu_b {\bf v}_{ab} \cdot \nabla W_{ab}.
\end{equation}

In the momentum and energy equations the quantities $\Pi_{ab}$ and $\Omega_{ab}$ are dissipation terms which we discuss below.

The equation of state used is that of the perfect fluid \begin{equation}P=(\Gamma-1)N\epsilon\end{equation} and is solved through a solving a nonlinear equation in the conserved variables, as described in \citet{chowandjoe}. Here $\Gamma$ is assumed to be constant, which allows the local sound speed, $c_s$ to be derived as

\begin{align}
c_s^2&=\frac{1}{h}(\frac{\partial P}{\partial n}+\frac{P}{n^2}\frac{\partial P}{\partial \epsilon})\notag\\
&=\frac{\Gamma(\Gamma-1)\epsilon}{1+\Gamma\epsilon}\end{align}

An issue which arises in the formulation of relativistic equations by using SPH particles is whether or not the kernels should take a Lorentz invariant form.  However, it is clear that the representation  of the gas by a set of SPH particles is not an invariant process. For example, in our frame we might begin with the SPH particles arranged on a cubic grid in analogy to the subdivision of space into cubical cells in a finite difference scheme.  In another inertial frame the SPH particles will appear to  be on a non-cubical grid, and the cells of the finite difference scheme will no longer be cubes. The computing observer in this other frame might decide to use either more or fewer particles and arrange them in some other way. The only issue is to calculate the dynamics as accurately as possible for given resources. 

Furthermore, we argue that the use of a flat space kernel is valid under the assumption that space is locally flat and that the radius of curvature of the background space time is large compared to the smoothing length. Should the scale of hydrodynamic variation (relative to the smoothing length) become small, particularly in an anisotropic fashion, then there is a clear argument for using spheroidal kernels \citep{fulbright95}
. The complications caused by this variation, in terms of relativistic consistency, are considered too great for this work and so are not attempted at this time.   

\section{Dissipation Terms}

A natural way to set up dissipation terms for SPH would be to use the continuum dissipation terms. In the case of non relativistic fluid dynamics it has turned out that more robust algorithms can be found which are not based directly on the continuum dissipation terms  (though they are equivalent when the continuum limit of the SPH equations is taken).  In the case of relativistic fluids the situation is more complicated because there are no satisfactory dissipation terms to begin with.  Those proposed by \citet{landau59},
\citet{eckart40} 
 or \citet{weinberg} 
are known to lead to instabilities (see \citet{israel76}, \citet{israel79} and \citet{hiscock},).

The dissipation terms were set up by \cite{chowandjoe} in analogy to the dissipation terms used in Riemann solver methods.  These latter dissipation terms are based on adding to the average flux from left and 
right cells a term of the form

\begin{equation}
\sum_d  | \lambda^d |  \left [ {\bf e}^d \cdot ({\bf F}_\ell- {\bf F}_r )\right ] {\bf e}_r
\end{equation}
where the $\lambda$ denote eigenvalues of the gas dynamic equations, the ${\bf e}$  denote eigenvectors and $r$ and $\ell$ denote right and left states.  These eigenvalues have the dimensions of velocity and represent signal velocities.  The ${\bf F}$ are the dependent variables of the differential equations: in the present case the relativistic momentum, energy and baryon number.  In addition various limiters may be used to prevent unwanted oscillations (see for example \citet{leveque}).

Following this idea the simplest dissipation term for the SPH momentum equation is
\begin{equation}
\Pi_{ab}=-\frac{Kv_{sig}(\mathbf{q}_a-\mathbf{q}_b) \centerdot \mathbf{j}}{\bar{N^*}_{ab}},
\end{equation}
where the eigenvalue is replaced by a signal velocity $v_{sig}$ which we discuss below, and the eigenvectors are replaced by 
\begin{equation}
\mathbf{j}=\frac{\mathbf{r}_{ab}}{|\mathbf{r}_{ab}|},
\end{equation}
which is the unit vector from particle $b$ to $a$.

The corresponding quantity for  the energy equation becomes
\begin{equation}
\mathbf{\Omega_{ab}}=-\frac{Kv_{sig}(e_a-e_b) \mathbf{j}}{\bar{N}^*_{ab}}
\end{equation}

In these expressions $\bar{N}^*_{ab}$ is an average of  $N^*_a$ and $N^*_b$ which in this paper we take to be $(N^*_a + N^*_b)/2$ and $K$ is a constant.  The quantity $v_{sig}$ denotes the signal velocity which is the analogue of the eigenvalues used in the Riemann solver methods. We have also symmetrised the dissipation terms to guarantee that linear and angular momentum will be conserved. In this paper the dissipation terms are set to zero if the particles are moving away from one another. 

Although the form of the dissipation terms given above are consistent with the global conservation of linear and angular momentum and energy they are not consistent with the requirement that viscous dissipation should result in an increase in the thermal energy.   Returning to (\ref{momdef}) and (\ref{energydef}) and taking the special relativistic limit, we can write the thermal energy $\epsilon$ in the form 
\begin{equation}
\epsilon = \gamma (  e - {\bf v} \cdot {\bf q}) -1,\label{therm}
\end{equation}
from which it follows that
\begin{equation}
\frac{d\epsilon}{dt} = \gamma \frac{d e}{dt} - \gamma {\bf v} \cdot \frac{d {\bf q}}{dt} - \frac{P}{N} \frac{d \gamma}{dt}.
\end{equation}
If $s$ denotes the entropy per mass in the co-moving frame we can write
\begin{equation}
T\frac{ds}{dt } = \frac{d\epsilon}{dt} - \frac{P}{n^2} \frac{dn}{dt},
\end{equation}
and substituting the non dissipative  momentum and energy equations  (\ref{momeqn}) and (\ref{eneqn}) together with (\ref{therm}) the change in entropy is zero as expected.  If we include the dissipation terms we find the rate of change of $\epsilon$ for SPH particle $a$ can be written
\begin{align}   
\frac{d\epsilon_a}{dt} = & - \frac{P}{N} \frac{d \gamma}{dt}  + \frac{\gamma_aP_a}{n_a^2} \sum_b m_b {\bf v}_{ab} \cdot \nabla _a W_{ab} \notag \\
  &+ \gamma_a \sum_b m_b \left (\Pi_{ab} {\bf v}_a 
                              - \Omega_{ab}  \right ) \cdot \nabla_a W_{ab}.
\end{align}
The first two terms on the right hand side are not associated with dissipation. The last term incorporates both viscous dissipation and thermal conduction.  The rate of change of the entropy per mass is then
\begin{equation}
T \frac{ds}{dt} =  \gamma_a \sum_b m_b \left (\Pi_{ab} {\bf v}_a 
                              - \Omega_{ab}  \right ) \cdot \nabla_a W_{ab}\label{e:tds}
\end{equation}
where the right hand side only involves the dissipation terms. In this expression for the entropy change $s$ is measured in the co-moving frame, while the time is measured in the laboratory frame. It is useful to write the rate of change  of entropy per unit mass $s$ as the four divergence of an entropy current $S^\mu = nsU^\mu$ according to
\begin{equation}
N\frac{ds}{dt} = \frac{\partial S^\mu}{\partial x^\mu}= \frac{\partial (Ns)}{\partial t} + \nabla \cdot (Ns {\bf v})\label{e:entropy}.
\end{equation}

It is convenient to first consider the viscous dissipation terms.  These can be isolated by setting the thermal terms to zero.  We find from (\ref{e:tds}) that the viscous dissipation term is
\begin{eqnarray}
\lefteqn{
 \gamma_a \sum_b m_b \frac{K v_{sig}}{\bar N_{ab}} }\notag\\
&& {} \left (\gamma_a -\gamma_b - (\gamma_a {\bf v}_a  -\gamma_b{\bf v}_b )\cdot {\bf j}({\bf v}_a \cdot {\bf j}) \right ) {\bf r}_{ab} F_{ab},
 \end{eqnarray}
where we have replaced $\nabla_a W_{ab}$ by ${\bf r}_{ab} F_{ab}$ where $F_{ab} \le 0$ is a function symmetric in the coordinates of particles $a$ and $b$.

This expression is not positive definite because the terms from $\Pi_{ab}$ involve velocities along the line joining the particles while $\gamma$ terms from the energy involve $v^2$.  To make the viscous dissipation positive definite we need an alternative form of $e$  in the dissipation terms. A simple and effective choice is the following. We first define  ${\bf v} \cdot {\bf j} =  v^*$, and $\gamma^*$ by $1/\sqrt{ 1- ( v^*)^2 }$. We then replace $e_a$ in the dissipation term by $e^*$ where
\begin{equation}
e^* = \gamma^*(1+\epsilon + \frac{P}{n} ) - \frac{P}{N}.
\end{equation}
the viscous dissipation term  then becomes
\begin{equation}
\gamma_a \sum_b m_b \frac{K v_{sig}}{\bar N_{ab}}\left( \sqrt{\frac{1-v^*_a}{1+v^*_a}} -v^*_a\sqrt{\frac{1-v^*_b}{1+v^*_b}} \right ) {\bf r}_{ab} F_{ab},
 \end{equation}
 To establish the sign of this expression we note with
 \begin{equation}
v^*_{ab} = \frac{v^*_a-v^*_b}{1- v^*_a v^*_b},
\end{equation}
we can form
\begin{equation}
\gamma^*_{ab} = \frac{1}{\sqrt{  1- (v^*_{ab})^2 } } = \gamma^*_a \gamma^*_b ( 1- v^*_a v^*_b)
\end{equation}
and the dissipation term can then be written
\begin{equation}
 \gamma_a \sum_b m_b \frac{K v_{sig}}{\gamma^*_a \bar N_{ab}}  \left ( 1- \gamma^*_{ab}  \right ) r_{ab} F_{ab}.
 \end{equation}
Since $\gamma^*_{ab} \ge 1$ and $F_{ab} \le 0$ the viscous dissipation term is positive definite.  The dissipation terms which we use in our equations are therefore $\Pi_{ab}$, which is unchanged, and
\begin{equation}
\Omega_{ab}=-\frac{Kv_{sig}(e^*_a-e^*_b) \mathbf{j}}{\bar{N}_{ab}}.
\end{equation}

The rate of change of entropy can be obtained from (\ref{e:entropy}) when the only dissipation is due to viscosity and from  the argument above, the change in the entropy from viscous dissipation is positive whether measured in the laboratory or the co-moving frame.  Unfortunately there is no agreed form of the transformations of the thermodynamic quantities between different inertial frames. Furthermore, although \citet{weinberg}, \citet{eckart40} and \citet{landau59} give entropy currents these depend on the forms they choose for the dissipation tensor.  None of the dissipation terms proposed in the literature are stable \citep{hiscock}. As a consequence it is not clear to us how to construct an appropriate entropy change from our dissipative terms when the thermal terms are included in the dissipation.

\subsection{Signal Speeds}
The signal speed $v_{sig}$ was considered by \citet{chowandjoe}.
 In the non relativistic case  $v_{sig}$ is the speed of approach of signals sent from SPH particles $a$ and $b$ towards each other \citep{joe92}.
  If these particles are separated by a distance $r$, and they are at rest,  the signals meet at a time $r/(c_a + c_b)$ where $c_a$ is the speed of sound of particle $a$. The signal speed is then taken as $c_a+ c_b$.  If the particles are moving terms must be added because the speed of sound is the propagation relative to the fluid which is itself moving.  In the non relativistic case the signal speed becomes 
\begin{equation}
v_{sig} = (c_a + c_b) - {\bf v_{ab}} \cdot {\bf j}.
\end{equation}

\citet{chowandjoe} tested several generalizations of this signal speed making allowance for the relativistic velocity addition formula.  However, they were not able to derive a suitable $v_{sig}$.

It is possible to be more precise about the form of $v_{sig}$ by considering the sound waves sent between two SPH particles $a$ and $b$.  The procedure we follow is to determine the wave number 4-vector of a sound wave from a particle $a$ towards particle $b$ in the co-moving frame of $a$. We then transform this back to our laboratory frame to get the apparent speed in the laboratory frame. We then repeat the procedure for particle $b$.  Finally we  construct  $v_{sig}$ by taking the average of these speeds.  

With this in mind, we take two inertial frames, the lab frame $K$, in which two particles $a$ and $b$ are moving, and the frame $ K^{\prime}$, in which particle $a$ is at rest. At some time particle $a$ emits a signal towards  the second particle $b$. In $K^{\prime}$, we know this signal to travel at the rest sound speed, $c_s$ which can be calculated by thermodynamic quantities known in the co-moving frame. Our aim is  to determine the apparent speed of sound $\hat{c}_s$ in the lab frame $K$.

Let the wave number 4-vector of the emitted sound  in the frame $K$ be $k^\mu = (\omega, {\bf k}) $ where $\omega = k \hat{c}_s$  and  ${\bf k}$ is directed from $a$ towards $b$. In the frame $K^\prime$ of particle $a$ the wave number 4-vector is  $(k^{\prime})^{ \mu } = (\omega^{\prime}, {\bf k}^{\prime}) $  where $\omega^{\prime} = k^{\prime} c_s$.  For convenience we rotate the laboratory axes so that the $x$ axis of the lab frame is parallel to ${\bf v}_a$ and denote the angle between the new  $x$ axis and the vector  ${\bf r}_{ba}$ from $a$ to $b$ by $\theta$.  

The Lorentz transformation between the frames $K$ and $K^{\prime}$ shows that 
\begin{equation}
\omega^{\prime} = \gamma_a \omega \left ( 1- \frac{v_a \cos{(\theta)} }{\hat{c}_s} \right ),
\end{equation}
together with  transformations of the wave number which lead to the aberration formula.  From the invariance of $k_\mu k^\mu$ we deduce
\begin{equation}
\omega^2 \left ( 1 - \frac{c^2}{\hat{c}^2_s} \right ) = \omega^{\prime2} \left (  1-   \frac{c^2}{c^2_s} \right ),
\end{equation}
where $c$ is the speed of light. In the following we will assume that the speed of sound is scaled with the speed of light as in our other equations.  Using the relation between the frequencies we can solve the previous equation for $\hat{c_s}$.   We find

\begin{equation}
\hat{c}_s=\frac{v_{||}(1-c_s^2)\pm c_s\sqrt{(1-v^2)[1-v_{||}^2-c_s^2v_{\perp}^2]}}{1-v^2c_s^2}\label{e:sigspeed}
\end{equation}

where  $v_{||}$ is the component of the ${\bf v}_a$ along the line joining the particles $a$ and $b$ and $\mathbf{v_\perp}$ is the perpendicular component (all measured in the laboratory frame). One can immediately see that for the one dimensional case (where $\mathbf{v}_\perp=0$), this reduces to the usual relativistic addition formula 
\begin{equation}
\hat{c}_s=\frac{c_s\pm v_{||}}{1\pm c_sv_{||}}
\end{equation}

Written in this way, equation (\ref{e:sigspeed}) for the signal speed is the same as the eigenvalue  $\lambda_\pm$ deduced by \citet{font94}
 for their acoustic waves travelling within a relativistic medium. By using this definition for our signal speeds (equation (\ref{e:sigspeed})) we ensure that the sound speeds remain causal and  the artificial viscosity terms ($\Pi_{ab}$ and $\Omega_{ab}$) do not lead to acausal communication.

\begin{figure}
\begin{center}
\includegraphics[width=\columnwidth]{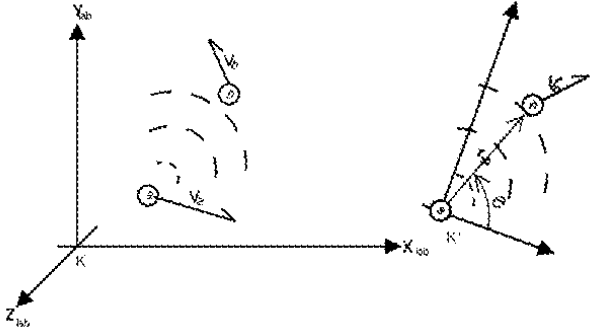}
\caption{Frames of Reference for soundspeeds: In the laboratory frame (on the left) we see two particles, $a$ and $b$, travelling with unique velocities when $a$ sends a signal to $b$. The right frame shows $a$ at the origin, with the x-axis aligned with $a$'s observed velocity in the lab frame. Particle $a$ still transmits to particle $b$, but at a different speed.}
\label{f:soundspeed}
\end{center}
\end{figure}

\section{TimeStep Controls}
The algorithm used for these tests is based upon the predictor-corrector integrator. In order to show how this method works, we introduce 3 vectors.
$\mathbf{F}$ is a five vector, containing the conserved quantities $N^*$, $e$ and $q^i$, $\mathbf{x}$ is the usual location vector and $\mathbf{H}$ holds all the hydrodynamic variables such as $\epsilon$, $P$, $\mathbf{v}$ and $\gamma$.
 We first predict $\mathbf{F}$ forward with an Euler step
\begin{equation}
\mathbf{F}=\mathbf{F}+\Delta t \frac{d\mathbf{F}}{dt}_{old}
\end{equation}

We then evolve the location vector forward through a second order prediction
\begin{equation}
\mathbf{x}=\mathbf{x}+\Delta t \mathbf{v} +\frac{1}{2}(\Delta t)^2\frac{d\mathbf{v}}{dt}\label{movexeqn}\end{equation}
For efficiency, we approximate the time derivative of the velocity with the known $\frac{d\mathbf{q}}{dt}$. Tests were also run replacing equation (\ref{movexeqn}) with a standard predictor-corrector which produced fundamentally the same results, indicating that this is a valid approximation.

With these predicted values of $\mathbf{F}$, we can solve for $\mathbf{H}$ using a Newton-Raphson root-finding method. The new $\mathbf{H}$ values allow us to generate new rates of change, $\frac{d\mathbf{F}}{dt}$. These are used, in turn, to correct $\mathbf{F}$
 through \begin{equation}
\mathbf{F}=\mathbf{F}+\frac{\Delta t}{2}( \frac{d\mathbf{F}}{dt}-\frac{d\mathbf{F}}{dt}_{old})\end{equation}

Finally, we can correct the values for $\mathbf{H}$ using the updated $\mathbf{F}$.

The timestep, $dt$, is given by
\begin{equation}
dt=\min\{C_vdt_v,C_adt_a,C_hdt_h\}\end{equation}
where $C_a, C_v,  C_h$ are constants and
\begin{align}
dt_v&=\min_i{\frac{h_i}{v_{sig}}_i}\\
dt_a&=\min_i{\sqrt{\frac{h_i}{|\mathbf{q}_i|}}}\\
dt_h&=\min_i{\frac{h_i}{\max_j{|{\mathbf{v}_i-\mathbf{v}_j}|}}}\end{align}
The first two of these are the usual Courant conditions on the velocity and the acceleration, respectively.
The third is that devised by \citet{thacker00} and is a new condition, based upon the smoothing radius divided by the highest relative velocity in that region. The algorithm seems rather insensitive to choice of the limiting constants and typical values chosen for $C_v$, $C_a$ and $C_h$ are 0.2, 0.35 and 0.35 respectively.

It should also be noted that \citet{thacker00} use the same `Courant' conditions, but use 0.4, 0.25 and 0.2 as their limiting parameters $C_v$, $C_a$, and $C_h$.  The dominant limiting factor is usually $dt_v$, and it has been further limited in our scheme in an attempt to reduce overshooting due to the frequent supersonic motions and large gradients involved.

\section{Numerical Tests}
\subsection{Boundary Conditions\label{s:boundarycond}}
The hydrodynamic test problems presented in this paper require a fixed box of known dimensions and so we employ a rectangular box. Two of the dimensions are periodic, and the third uses ghost-particles to maintain its structure. Periodicity is maintained by simply mapping the leftmost particles into pseudo-positions against the right boundary (and vice versa), and modifying the location vectors accordingly.

The computational box used is defined as $-0.5\leqslant x\lesssim 0.5$ in the x-direction, and by the number of particles and the particle spacing in both the y and z directions. The initialisation procedure involves laying down the grid from $x=-0.5$ until close to the origin, where the particle spacing smoothly changes across to the lower density values in the right hand region. The upper x-boundary is then chosen to fit the particle spacing in exactly at around $x=+0.5.$

The smoothing procedure involves smearing any discontinuity out across a few particle spacings in the $x$ direction. To accomplish this, we simply define the particle position to be in one of three zones, the left steady state ($A_L$), right steady state ($A_R$), or a transition region in between. The field variables $A$ are then smoothed over the range by the rule

\begin{equation}
A(x)=\frac{A_L+A_R e^{\frac{x}{\Delta x}}}{1+e^{\frac{x}{\Delta x}}}\end{equation}

where $\Delta x$ is the particle separation chosen to maintain the relation between particles and desired density distribution. The particles are all given the same baryon number $\nu_b$. The initial (high density) particle spacing is chosen as an input parameter. This is then varied as the particles are initially positioned, maintaining the relationship between $\Delta x$ and $\nu_b$ to be the local baryon number density ($N^*$).
 This is the same method used by \citet{chowandjoe}
 in their 1-D application where it performed satisfactorily.

\section{Shock Tube Results}
\subsection{Newtonian Shocks}
It is important that any new relativistic algorithm is first tested to ensure it generates the correct non-relativistic limit. The shock problem of \citet{sod} is  probably the most recognised and so we have reproduced it here as a one dimensional problem, modelled in full 3-D. The initial configuration involves a discontinuity at $x=0$ separating two states, the high energy left state ($N^*_l=1\times 10^5,\epsilon_l=2.5\times 10^{-5}$ and the right state ($N^*_r=0.125\times 10^5,\epsilon_r=2.0\times 10^{-5}$).   
 This configuration was initially done by \citet{siegler99} and is replicated here, although calculated as a full 3-dimensional simulation. The numbers of particles in each coordinate direction ($\text{Num}_i$) are given as (($\text{Num}_x=2000,\text{Num}_y=8,\text{Num}_z=8$):($\text{Num}_x=1000,\text{Num}_y=4,\text{Num}_z=4$)) which corresponds to a computational box of $1.049\times 0.004\times 0.004$. The smoothing lengths are chosen and evolved to maintain $\sim57$ neighbours.

All of the following calculations are performed in the special relativistic limit, where $\alpha=1$ and $\eta^{ij}=\eta_{ij}=\text{diag}\{1,1,1\}.$

\begin{figure}
\begin{turn}{0}\includegraphics[width=\columnwidth]{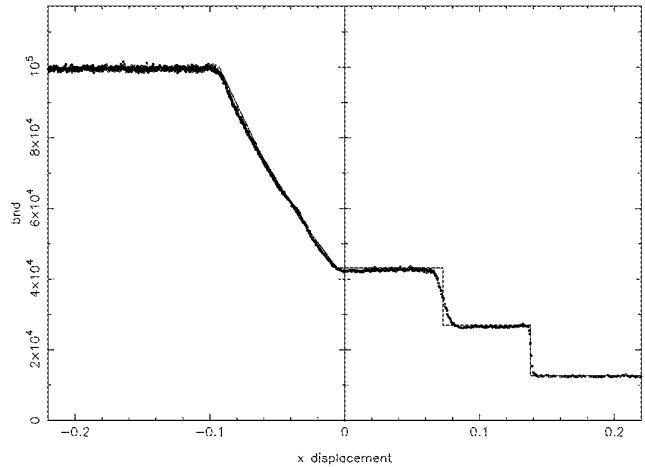}\end{turn}
\caption{3D simulation of the Newtonian Shock showing baryon number distribution ($D$) \label{f:3dsod1}}
\end{figure}

\begin{figure}
\begin{turn}{0}\includegraphics[width=\columnwidth]{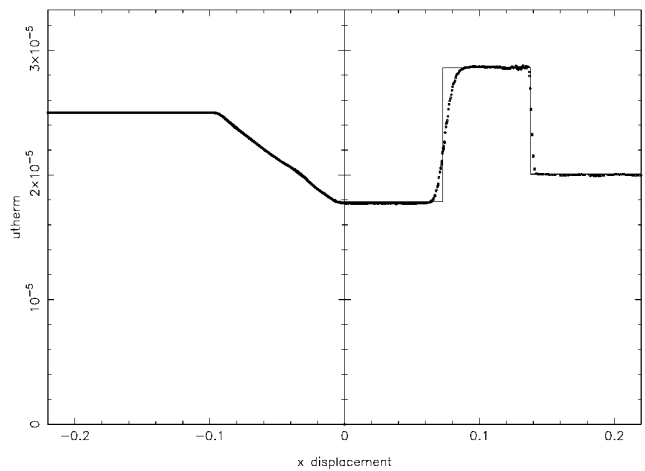}\end{turn}
\caption{3D simulation of the Newtonian Shock showing thermal energy, $\epsilon$\label{f:3dsod3}}
\end{figure}

\begin{figure}
\begin{turn}{0}\includegraphics[width=\columnwidth]{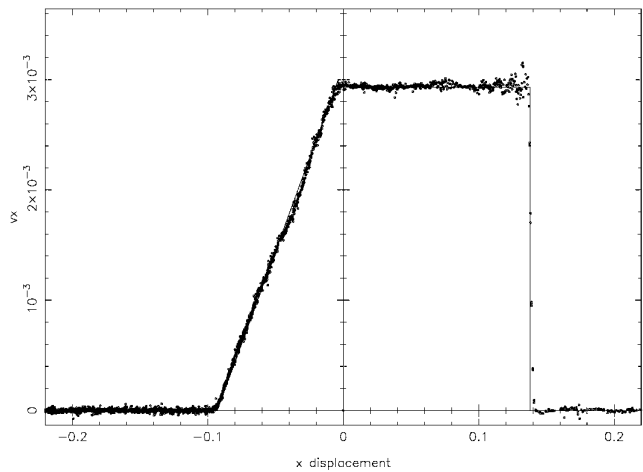}\end{turn}
\caption{3D simulation of the Newtonian Shock showing velocity in the x-direction\label{f:3dsod4}}
\end{figure}

\begin{figure}
\begin{turn}{0}\includegraphics[width=\columnwidth]{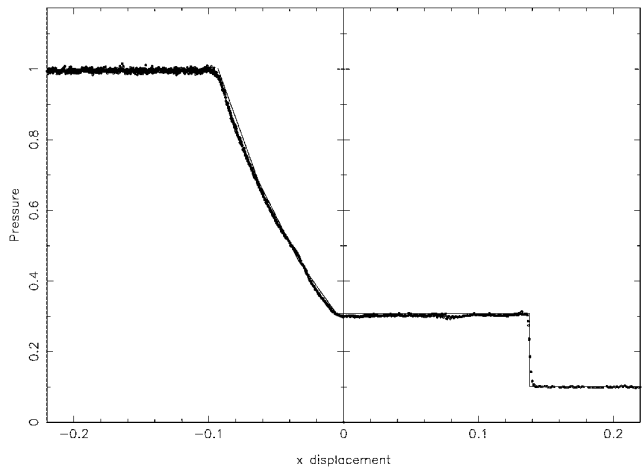}\end{turn}
\caption{3D simulation of the Newtonian Shock showing isotropic pressure\label{f:3dsod2}}
\end{figure}

Four graphs are shown, the number density, the isotropic pressure, the thermal energy and the velocity in the x-direction (the direction of shock propagation). Note the diffusion evident across the contact discontinuity in both Figure \ref{f:3dsod1} and \ref{f:3dsod3}. This can be reduced by increasing the resolution of the simulation, which is currently limited by the constraints of single processor machines. In this particular example, the contact discontinuity is resolved over $\sim 35$ particle spacings. These correspond to a computational time of $\sim25.$ This time corresponds to $\sim26$ sound crossing times which accounts for the scatter particularly evident in Figure \ref{f:3dsod4}. Figure \ref{f:3dsod2} shows some oscillation across the contact discontinuity, but no spiking as evidenced in other SPH discretizations. The oscillation can be greatly reduced by tuning the viscosity parameter $K$, which in this case is 1.4, but this was not pursued to any great extent.


\subsection{Relativistic Shocks}
The next example we will show is the shock tube studied in detail in one-dimension by \citet{hawley84}, \citet{schneider93}, \citet{marti96} and \citet{siegler99}
 . Again the higher density and pressure state is to the left, and lower energy states to the right. When the two states are brought together, a shock wave propagates into the low density gas to the right, and a rarefaction fan propagates leftwards into the higher density state. Between these two waves, there is a contact discontinuity where the pressure and velocity are continuous, but the density drops. The initial conditions across the discontinuity are pressure; $13.3:1.0\times10^{-6}$ and Number Density; 10:1.

\begin{figure}
\begin{turn}{0}\includegraphics[width=\columnwidth]{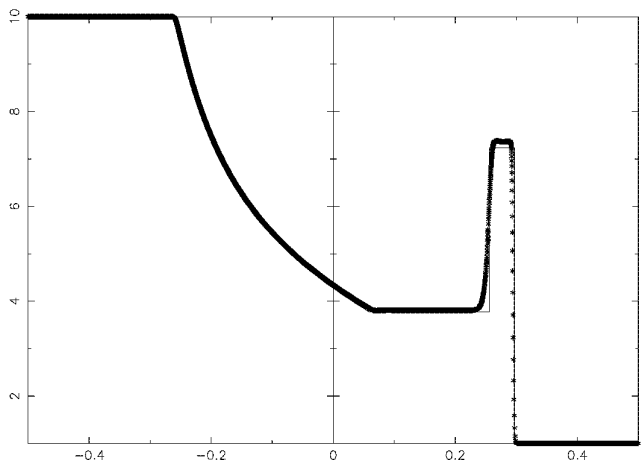}\end{turn}
\caption{1D simulation of the Relativistic Shock showing Baryon Number Density distribution\label{f:hires1dshockD}}
\end{figure}

\begin{figure}
\begin{turn}{0}\includegraphics[width=\columnwidth]{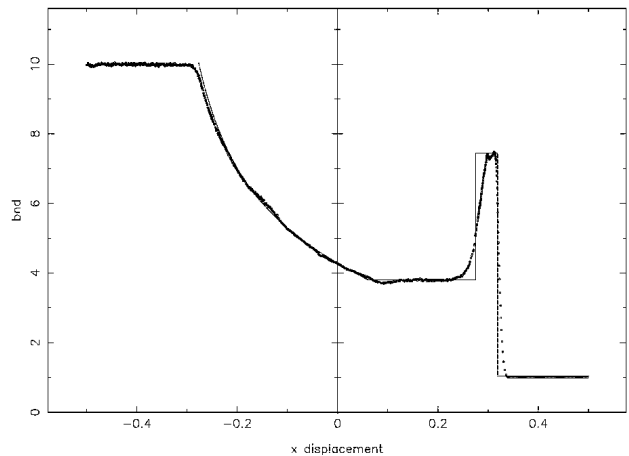}\end{turn}
\caption{3D simulation of the Relativistic Shock showing Baryon Number Density  Distribution\label{f:3dBND}}
\end{figure}

The baryon number density distribution is shown in Figure \ref{f:hires1dshockD} where one can see that the locations of the shock front, contact discontinuity and the head and tail of the rarefaction fan are well resolved. There is a slight over-estimate of the magnitude of the shocked shell. This is compared to Figure \ref{f:3dBND} which shows the identical 1-Dimensional phenomena, although the calculation is performed in 3 spatial dimensions. Note that with the new $v_{sig}$ and viscosity terms, there is no spike exhibited at the contact discontinuity.
 
In moving from one to three dimensions there is significant increase in demand placed upon the dissipation terms. There is a lot more opportunity with the added degrees of freedom for the particles to interpenetrate, and for false  communication to occur. Therefore we would not expect that what occurs in 1D necessarily carries over into 3D. This is shown in the lack of resolution of the contact discontinutity and smearing of the shockfront. There is also a small amount of diffusion evident in the tail of the rarefaction fan. This is due to the artificial viscosity terms being unable to take the directionality of the expansion into account. As a consequence, the expansion in x corresponds to particles also having y and z velocities. Because of the fixed boundaries these velocities correspond to colliding orbits in the y and z plane, which erroneously activate the artificial viscosity.
The post-shock ringing is also seen in the dense shell.

\begin{figure}\begin{center}
\begin{turn}{0}\includegraphics[width=\columnwidth]{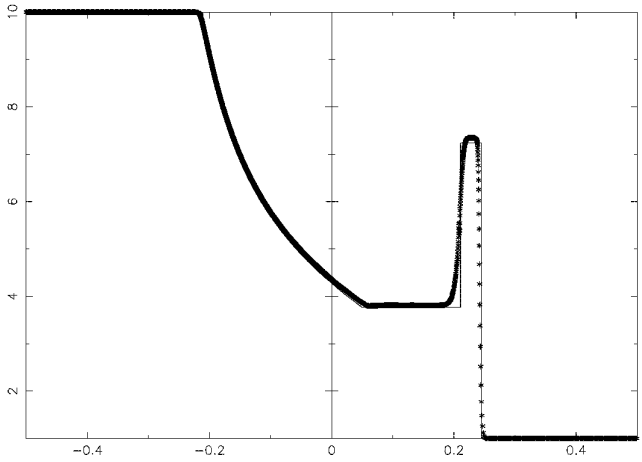}\end{turn}
\caption{1-D baryon number density versus x location, with identical particle resolution in the $x$ direction as Figure \ref{f:3dBND} \label{f:3dcompare1}}\end{center}
\end{figure}

Many of these effects can be seen to be diminished by increasing the number of particles in the simulation. Due to the constraints of memory, we are currently unable to perform the 3-dimensional calculation at the same x-resolution as the example shown in Figure \ref{f:hires1dshockD}. To highlight the issues of particle numbers, we then show Figure \ref{f:3dcompare1}, which corresponds to a 1-dimensional calculation with identical x resolution as the Figure \ref{f:3dBND}. (350 particles in the x-direction across low density region). Both Figure \ref{f:3dBND} and \ref{f:3dcompare1} show the simulation after 2000 timesteps. In the 1-dimensional calculation, the rarefaction fan is much better resolved, due to the viscosity terms not being activated. Notice also how there is no ringing in the shell region. The contact discontinuity is also better resolved, although less sharp than the high resolution Figure  \ref{f:hires1dshockD} which also displays a flatter dense shell plateau. 

\begin{figure}
\begin{turn}{0}\includegraphics[width=\columnwidth]{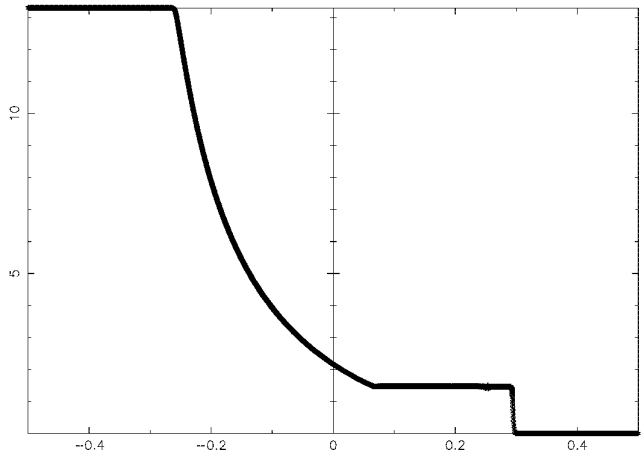}\end{turn}
\caption{1D simulation of the Relativistic Shock showing Pressure Distribution\label{f:hires1dshockP}}
\end{figure}

\begin{figure}
\begin{turn}{0}\includegraphics[width=\columnwidth]{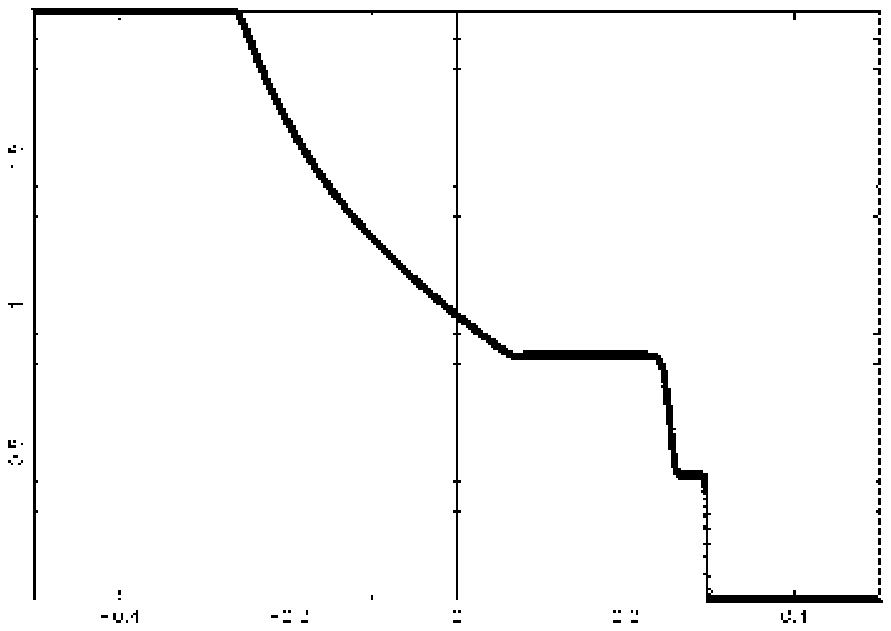}\end{turn}
\caption{1D simulation of the Relativistic Shock showing Thermal Energy  Distribution\label{f:hires1dshockU}}
\end{figure}

\begin{figure}
\begin{turn}{0}\includegraphics[width=\columnwidth]{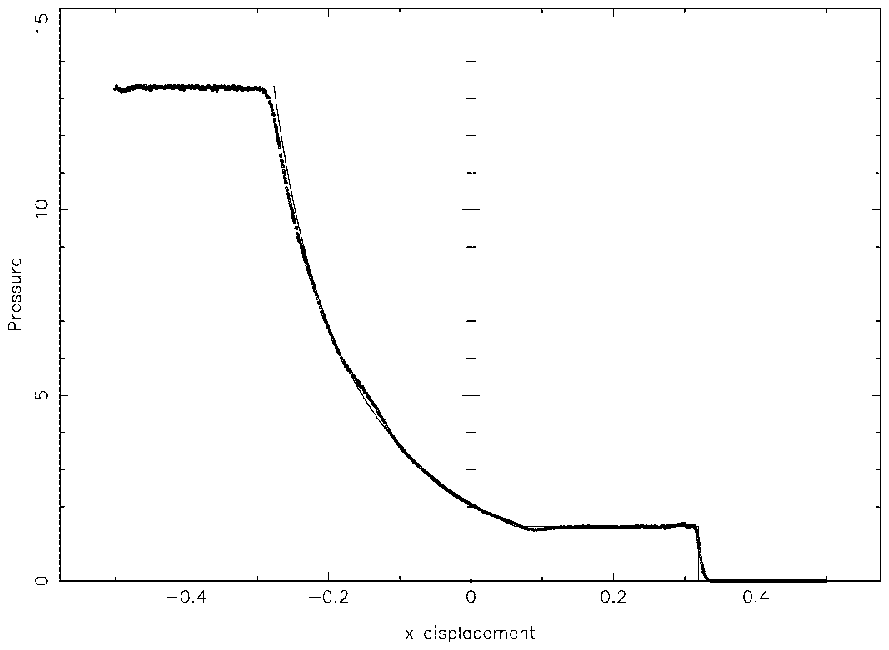}\end{turn}
\caption{3D simulation of the Relativistic Shock showing Pressure Distribution\label{f:3dP}}
\end{figure}

\begin{figure}
\begin{turn}{0}\includegraphics[width=\columnwidth]{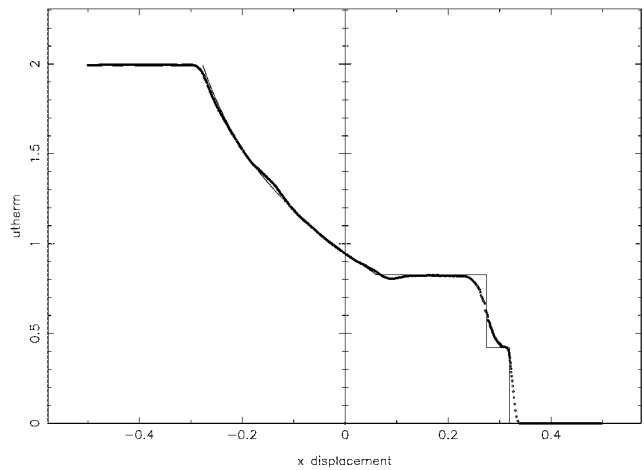}\end{turn}
\caption{3D simulation of the Relativistic Shock showing Thermal Energy  Distribution\label{f:3dU}}
\end{figure}

In order to examine the performance of the diffusion, we draw your attention to the 1-dimensional calculations shown in Figures \ref{f:hires1dshockP} and \ref{f:hires1dshockU} and the three dimensional simulations shown in Figures \ref{f:3dP} and \ref{f:3dU}.


In the one dimensional calculation, one can see the sharp resolution of the shock, and diffusion on the tail of the contact discontinuity. Also clear is the scatter in the pressure distribution at the same location. This spike is much curtailed in comparison to models which integrate the thermal energy equation rather than the total energy (see \citet{siegler99}).
 In 3 dimensions this spike is reduced to the same level as the surrounding scatter, although the diffusion caused by the large amounts of viscosity required to stabilise the post-shock region is clear, highlighted by the excessive smearing of the contact discontinuity in Figure \ref{f:3dU}.

\begin{figure}
\begin{turn}{0}\includegraphics[width=\columnwidth]{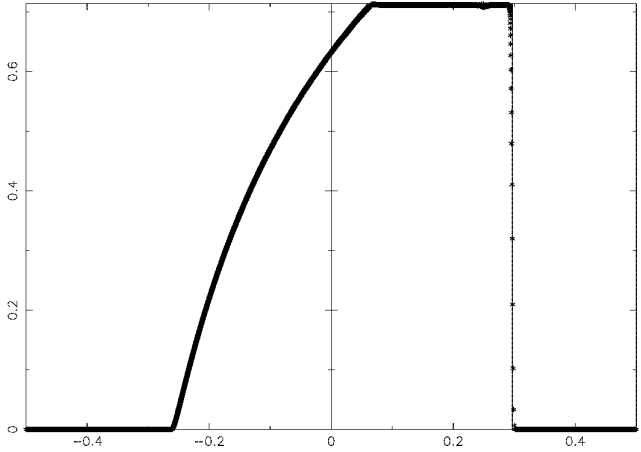}\end{turn}
\caption{1D simulation of the Relativistic Shock showing Velocity Distribution\label{f:hires1dshockV}}
\end{figure}

\begin{figure}
\begin{turn}{0}\includegraphics[width=\columnwidth]{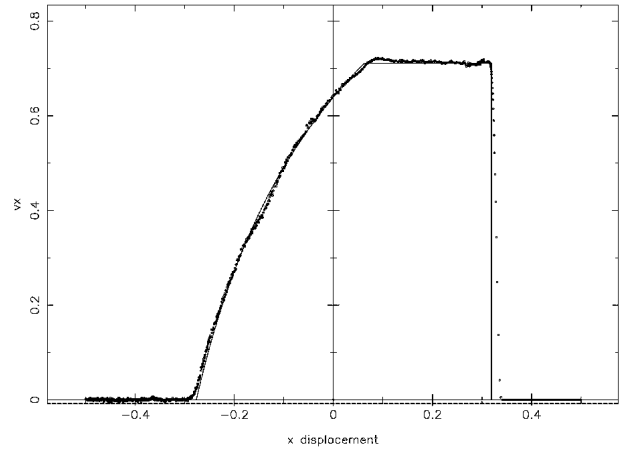}\end{turn}
\caption{3D simulation of the Relativistic Shock showing Velocity Distribution\label{f:3dVx}}
\end{figure}

The curves most sensitive to scatter are the velocity curves, shown in Figure \ref{f:hires1dshockV} and \ref{f:3dVx} (and \ref{f:3dsod4}). This is due to the manner in which the velocity components are calculated. This involves a rootfinder for the pressure term which then allows the magnitude of the velocity to be deduced. The magnitude of the velocity can  then be decomposed into its constituent components through a comparison to the conserved quantity of momentum. In the one dimensional case, this is a simple deconvolution, but is prone to more scatter in higher dimensions, as seen in Figure \ref{f:3dVx}

\begin{figure}\begin{center}
\begin{turn}{0}\includegraphics[width=\columnwidth]{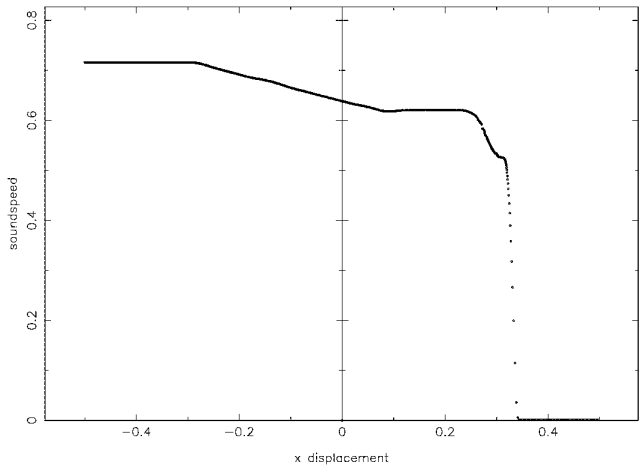}\end{turn}
\caption{Sound Speed\label{f:3dcs}}\end{center}
\end{figure}

The amount of scatter evident is due to the number of sound crossing times required to run the model. As pointed out previously, the computational box is defined by the number of particles, not a physical distance. In this particular case, for this resolution this works out to be a box with dimensions (1.05x0.02x0.02). If we look at Figure \ref{f:3dcs} we can see that the sound speed is a high 0.72$c$ in the high density region. For the time of 0.3865 units that the previous figures are all produced at, we can see that this corresponds to nearly 15 sound crossing times.

\begin{figure}\begin{center}
\begin{turn}{0}\includegraphics[width=\columnwidth]{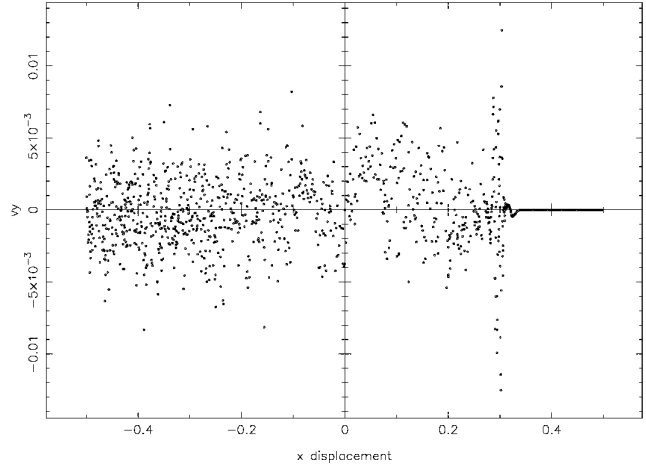}\end{turn}
\caption{Velocities transverse to the direction of shock propagation\label{f:3dvy}}\end{center}
\end{figure}

Continuing this analysis further is the Figure \ref{f:3dvy} which shows the transverse velocities of the particles as a function of their x location. Note that these velocities are all under 1.5 per cent of the local sound speed.

\section{Summary}
In summary, we have shown here a set of  relativistic SPH equations which correspond to a static background potential (space-time). They have no destabilizing time derivatives of hydrodynamic variables on the right hand side  and have been shown to adequately capture basic shock flows.  The shocks are captured through a new signal velocity term, which avoids the restricting requirement of an exact Riemann solution needed by some applications. The new derivation of the artificial viscosity precludes any acausal communication between particles and avoids the false spiking of other implementations. This false shock should be avoided as it may cause erroneous source terms in simulations where nuclear interactions, such as burning, occur.

In their current guise the equations assume a stationary and static metric. The extension to metrics such as the Kerr is being studied.
 

\bsp

\label{lastpage}

\end{document}